\documentclass[twocolumn,showpacs,preprintnumbers,amsmath,amssymb]{revtex4}

\usepackage{graphicx}
\usepackage{dcolumn}
\usepackage{bm}
\usepackage{epsfig}
\usepackage{epstopdf}

\begin{document}


\title{Observation of ultra-broadband, beamlike parametric downconversion}

\author{Kevin A. O'Donnell and Alfred B. U'Ren}

\address{
Divisi{\'o}n de F{\' \i}sica Aplicada,  Centro de Investigaci\'{o}n Cient{\' \i}fica y de Educaci\'{o}n Superior de Ensenada \\
Apartado Postal 2732, Ensenada, Baja California, 22800 M\'{e}xico}

\date{\today}

%
\newcommand{\epsfg}[2]{\centerline{\scalebox{#2}{\epsfbox{#1}}}}

\begin{abstract}
We report spontaneous parametric downconversion having an unusually wide spectral bandwidth.
A collinear type-1 phase-matching configuration is employed with degeneracy
near the zero group velocity dispersion frequency.   With a spectral width of $1080 \, \mathrm{nm}$ and degenerate wavelength $1885 \, \mathrm{nm}$, the source also emits a high flux of $3.4 \! \times \! 10^{11} \, \mbox{s}^{-1} \mbox{W}^{-1}$ photon pairs constrained to a cone of only $\approx \! 2^\circ$ half-angle.  A rigorous theoretical approach is developed that confirms the experimental observations.  The source properties are consistent with an ultra-short photon-pair correlation time and, for a narrowband pump, extremely high-dimensional spectral entanglement.
\end{abstract}

\maketitle


The process of spontaneous parametric downconversion (SPDC), in
which pump photons split into pairs of signal and idler photons,
can exhibit a wide range of emission characteristics. In particular, the
spectral properties vary widely in practice and are intimately tied
to the phase-matching configuration in the second-order nonlinear crystal employed.
The purpose of this Letter is to demonstrate that, when certain conditions related to
phase matching are met, even a narrowband pump can
produce downconversion with an exceptionally large bandwidth. 
These same conditions also lead to 
a high flux of downconverted photons emitted into a narrow,
beam-like angular distribution.

The significance of broadband downconversion has been discussed previously.
A large bandwidth is consistent with a short correlation time
between the signal and idler photons of a given
pair\cite{strekalov05}. This time represents the resolution limit in
metrology applications, such as quantum optical coherence
tomography\cite{abouraddy03}, that rely on signal/idler arrival-time
differences. Also, for a narrowband pump, a wide downconversion bandwidth
creates a two-photon state having many Schmidt mode pairs or, equivalently, 
high-dimensional spectral entanglement\cite{zhang06a}. Many modes implies a large mutual
information, which is a measure of the information that can, in
principle, be shared by two parties through spectral photon pair
correlations\cite{zhang06b}. Further, Schmidt modes can be occupied
independently, so that having many available modes means that
individual photon pair behavior can be preserved at high light levels. Such 
behavior has been demonstrated with a
two-photon process that is linear instead of quadratic in its
power-dependence\cite{dayan05}.

There have been previous experimental demonstrations of broadband
SPDC, where bandwidths of tens of nm are not uncommon. For example,
in Ref.~\onlinecite{dayan05} a bandwidth (defined throughout as full
width at half maximum) of $31 \, \mathrm{nm}$ was reported for periodically-poled bulk 
KTP, while Ref.~\onlinecite{uren04} obtained a bandwidth of  
$\approx \! 40 \, \mathrm{nm}$ with a KTP waveguide, with both results being in the near
infrared. Elsewhere, a downconversion bandwidth of $\approx \! 115
\, \mathrm{nm}$ was reported near a mean wavelength of $840
\,\mathrm{nm}$\cite{nasr05}. There, the frequency spread of a
broadband pump was mapped into a yet wider range of downconverted
frequencies. A different approach was taken in
Ref.~\onlinecite{carrasco06}, which reported a bandwidth of $185 \,
\mathrm{nm}$ at center wavelength $812 \, \mathrm{nm}$. In this case the pump light was narrowband but tightly-focused; the
bandwidth arose from phase matching the wide angular range of
pump wave vectors.

In contrast, here we develop a broadband SPDC source with a narrowband, weakly-focused pump.  In a periodically-poled crystal, energy conservation and perfect phase matching imply that\cite{Fejer}
\begin{align}
 & \qquad \qquad \omega_p - \omega_s - \omega_i = 0 \quad \mathrm{and}  \label{freq}  \\
 & \Delta \mathbf{k} \equiv \mathbf{k}_p(\omega_p) -  \mathbf{k}_s(\omega_s) - \mathbf{k}_i(\omega_i) - \mathbf{k}_g = 0 \;  \label{vec} ,
\end{align}
where the subscripts $p$, $s$, and $i$ denote quantities associated
with, respectively, pump, signal, and idler photons.  Here, $\Delta
\mathbf{k}$ is the wave vector mismatch, $\omega_j$ ($j \! = \! p,
s, \mathrm{or} \, i$) is the frequency, $\mathbf{k}_j(\omega_j)$ is
the wave vector within the crystal, and $\mathbf{k}_g$ is the poling
wave vector.

To produce wideband phase matching, consider the
case when all vectors of Eq.~(\ref{vec}) are collinear and
are replaced by corresponding scalars.  We assume a
monochromatic pump with fixed $\omega_p$, but allow $\omega_s$ and $\omega_i$ to vary
from the degenerate frequency $\omega_d \equiv \omega_p/2$ as
$\omega_s \! = \! \omega_d + \delta \omega_s$ and  $\omega_i \! = \!
\omega_d + \delta \omega_i$.  Expanding $\Delta k$ into a power
series in $\delta \omega_s$ and $\delta \omega_i$, it follows that
\begin{equation}
  \Delta k =   {\sum_{n=1}^\infty} \; [ (-1)^{n+1} k_i^{(n)} (\omega_d) -  k_s^{(n)} (\omega_d) ] \; \delta {\omega_s}^n / n!\; ,
\label{delk}
\end{equation}
where the superscripts $(n)$ denote $n$th frequency derivatives,
$k_g$ has been chosen to provide perfect phase matching at
degeneracy, and it has been noted that Eq.~(\ref{freq}) implies that
$\delta \omega_i = -\delta \omega_s$.  In what follows, we broaden
the bandwidth over which $| \Delta k |$ is small by removing terms of
Eq.~(\ref{delk}) of increasing order in $\delta \omega_s$.  First,
for a type 1 interaction, $k_s(\omega_s)$ and $k_i(\omega_i)$ are
associated with the same crystal index, so all terms having odd $n$
in Eq.~(\ref{delk}) vanish.  Then, the $n \! = \! 2$ term of
Eq.~(\ref{delk}) may be eliminated by choosing $\omega_d$ so that
$k_{s,i}^{(2)} (\omega_d) \! = \! 0$ for the particular crystal, which is the condition of zero group velocity dispersion.  The series does not then begin
until the \textit{fourth} order in $\delta \omega_s$, implying a wide emitting bandwidth. 
It is notable that these conditions can only occur at degeneracy; otherwise, the $n$th term of Eq.~(\ref{delk}) contains $k_s^{(n)}$
and $k_i^{(n)}$ evaluated at different frequencies, and term cancellations for $n \! = \! 1, 2,$ and 3 are not possible.

We employ periodically-poled lithium niobate, with pump, signal, and
idler waves polarized along the $z$-axis and propagation along the
$x$-axis of the crystal.  Using a Sellmeier equation for the
extraordinary refractive index\cite{Jundt}, it is straightforward to
calculate the temperature-dependent $\omega_d$ for which
$k_{s,i}^{(2)} (\omega_d) \! = \! 0$.  At $20^\circ \mathrm{C}$,
this condition corresponds to a pump wavelength $\hat{\lambda}_p \!
= \! 2 \pi c / \omega_p \! = \! 958.7 \, \mathrm{nm}$, which in turn
implies a poling period $\hat{\Lambda}_g \! = \! 2 \pi / k_g \! = \!
28.1 \, \mu \mathrm{m}$ from Eq.~(\ref{vec}) at degeneracy.  Thus,
cancellation of all terms of Eq.~(\ref{delk}) for $n \! < \! 4$
is consistent with reasonable experimental parameters.

In experiments, the pump was a narrowband, continuous-wave,
Ti:sapphire laser having $\approx$1~W power\cite{Hansch} with a
birefringent filter added to tune over $910 \! - \! 980 \,
\mathrm{nm}$.  As indicated in Fig.~1, lenses focused the output
beam to a wide $110 \, \mu \mathrm{m}$ waist inside the $L \! = \! 1 \,
\mathrm{cm}$ long crystal, which had poling periods $\Lambda_g$ in
steps between $25.5 \, \mu \mathrm{m}$ and $28.7 \, \mu \mathrm{m}$
at 8 positions along its entrance face.  The light exiting the
crystal was incident on a Si Brewster plate that absorbed the pump
but transmitted the downconverted flux.  This flux then passed through a
limiting aperture and was focused by a $\mathrm{CaF}_2$ lens into a
spectrometer, which employed a highly efficient Littrow grating and
a sensitive cooled InSb detector.  The detector signal was processed
by a lock-in amplifier.

\begin{figure}[b]
  \epsfg{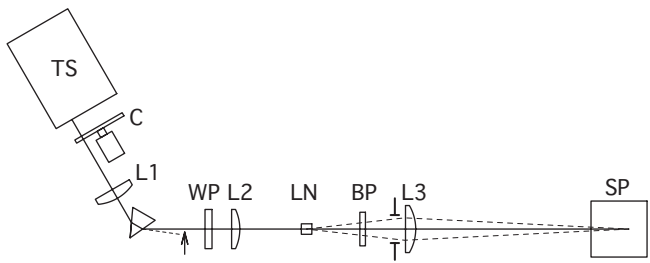}{.44} \caption{Broadband downconversion experiment.  The Ti:sapphire laser
(TS) beam is chopped (C) and focused into a lithium niobate crystal
LN by lenses L1 and L2. A half-wave plate WP rotates the pump
polarization.   A Si Brewster plate (BP, with plate normal directed
out of figure plane) isolates the downconversion, and the emission
half-angle subtended by the limiting aperture after BP defines
$\Delta \theta$.  The $\mathrm{CaF}_2$ lens L3 focuses the
light into the spectrometer SP.}
\end{figure}

It was verified that if either the pump or detected polarization was
rotated by $90^\circ$, the detector signal fell to
low levels as expected.  Even at room temperature, there was no
evidence of photorefractive crystal damage.  It was also found that
temperature-tuning the crystal could flood the spectrometer with
blackbody radiation but only weakly affected the SPDC spectrum, as
is consistent with broadband phase matching.  Thus the crystal was
instead operated at room temperature and the spectrum was
width-optimized by varying  $\lambda_p$ and $\Lambda_g$.  Indeed,
$\lambda_p$ and $\Lambda_g$ as will be quoted in the experimental
results differ slightly from $\hat{\lambda}_p$ and $\hat{\Lambda}_g$
calculated earlier.  These differences may be attributed to
optimization of phase matching throughout the limiting
aperture of Fig.~1.

With $\lambda_p \! = \! 942.5 \, \mathrm{nm}$ and $\Lambda_g \! = \!
27.4 \, \mu \mathrm{m}$, Fig.~2(a) shows measured spectra  for
several values of the limiting aperture angular radius $\Delta
\theta$.  For small $\Delta \theta$, the spectrum rapidly rises with
increasing $\Delta \theta$, while the curves for $\Delta \theta \! =
\! 1.5 ^\circ$, $2.0^\circ$, and $2.5^\circ$ are of similar height.
Most significantly, the spectra are remarkably broad.  
For the three larger $\Delta \theta$ of Fig.~2(a) the frequency bandwidth is 
$0.53 \, \bar{\omega}$, where $\bar{\omega}$ is the average frequency;
in terms of wavelength, this corresponds to an unprecedented $1080 \, \mathrm{nm}$
bandwidth.  The spectra are normalized in units of photon rate per unit frequency
per watt of pump power.  Integration of the spectrum
for $\Delta \theta \! = \! 2.5 ^\circ$ implies a photon pair
rate of $3.4  \times  10^{11} \mathrm{s}^{-1}
\mathrm{W}^{-1}$, or a SPDC power of $3.6 \times 10^{-8}$ times
the incident power.  These values (and, for that matter, the
spectra) are internal quantities within the nonlinear crystal, which
are inferred by applying calculated losses to external measurements.

\begin{figure}[b]
  \epsfg{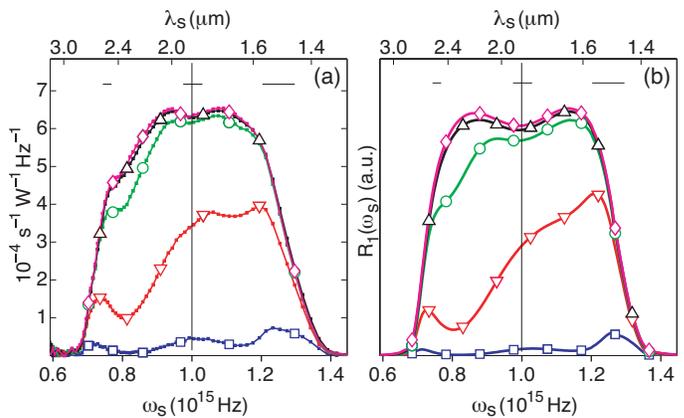}{.5} \caption{ (a)
Experimental downconversion spectra for a $1$ cm lithium
niobate crystal with poling period $27.4 \, \mu \mathrm{m}$, pumped
at wavelength  $942.5 \, \mathrm{nm}$.  Cases show collection angle
$\Delta \theta \! = \! 0.5^\circ$ (squares), $\Delta\theta \! = \! 1.0^\circ$ (inverted triangles), 
$\Delta \theta \! = \!1.5^\circ$ (circles), $\Delta \theta \! = \! 2.0^\circ$ (triangles),
$\Delta \theta \! = \! 2.5^\circ$ (diamonds).  (b) Corresponding theoretical
spectra for $942.5 \, \mathrm{nm}$, convolved over the spectrometer slit integration width.
Horizontal lines denote these $\omega_s$-dependent widths; vertical lines denote the
degenerate frequency. }
\end{figure}

In other data, we have found that for shorter $\lambda_p$ the SPDC
power falls; for $\lambda_p \! \le \! 939 \, \mathrm{nm}$ the
spectra fall to $ \le \! 10 \%$ of the levels of Fig.~2(a).  For
$\lambda_p \! > \! 942.5 \, \mathrm{nm}$ there is still significant
SPDC power, but the spectra sag for frequencies near the center of
Fig.~2(a).  Thus, while results for other $\lambda_p$ are not shown
here, the results of Fig.~2(a) with $\lambda_p \! = \! 942.5 \, \mathrm{nm}$ 
may be considered near-optimal.

To calculate the SPDC photon number spectrum $R_1(\omega_s)$, 
we develop a theoretical approach based on Ref.~\onlinecite{ou89}. It
follows that
\begin{equation}
R_1({\omega_s})=\frac{\epsilon_0}{2 \, \delta t \, \hbar \omega_s} \int dx'
\int dy' \int\limits_{z_0-c \delta t}^{z_0} dz' \langle
\hat{I}(\mathbf{r'},\frac{z_0}{c})\rangle
\end{equation}
where $\hat{I}(\mathbf{r},t) \! \equiv \! \hat{E}^{(-)}(\mathbf{r},t)
\hat{E}^{(+)}(\mathbf{r},t)$ is the intensity operator, 
$\langle \cdot \rangle$ indicates an expectation value with respect to the two-photon state,
$z_0$ is the distance along the propagation axis from the crystal to the detector,
$\mathbf{r'}$ refers to coordinates outside the crystal,
and $c \, \delta t$ is the length of the volume detected in time $\delta t$.
In the following $\theta_{s,i}$ and $\phi_{s,i}$ denote, respectively, the polar and
azimuthal internal propagation angles of the signal/idler waves.
For a Gaussian pump beam with waist $w_0$ at the crystal midpoint,
$R_1(\omega_s)$ may be written as
\begin{eqnarray}
&R_{1}(\omega_s) \propto  \frac{\omega_s
k_s^2 k_s^{(1)}}{n_s^2} \int\limits_{0}^{\pi}d
\theta_s \, \sin \theta_s \int\limits_{0}^{2\pi} d \phi_s \int d^3  k_i
\, (\omega_i / n_i^2) \nonumber
\\
&\times \, 
\eta_r(\theta_s,\phi_s)\, |g(\omega_s,\theta_s,\phi_s;\omega_i,\theta_i,\phi_i)|^2 \, \delta(\omega_p \! - \! \omega_s \! - \! \omega_i)
\label{E:R1}
\end{eqnarray}
where the $\omega_{s,i}$-dependence of $k_{s,i}$ is now implicit, 
$d^3k_i \! = \! k_i^2\,  k_i^{(1)} \sin \theta_i \, d\omega_i\, d\theta_i \,d\phi_i$,
and $\eta_r(\theta_s,\phi_s)$ is of unit height only within the limiting aperture (the physical aperture function $\eta(\theta_s,\phi_s)$ is mapped to the internal $\eta_r(\theta_s,\phi_s)$ by refraction at the crystal exit face).  The function $g$
represents the joint amplitude\cite{uren03}
\begin{equation}
g(\omega_s,\theta_s,\phi_s;\omega_i,\theta_i,\phi_i)= \exp{[-(k_\bot w_0 / 2)^2]} \, \frac{ \mbox{sin}(\beta \, L)}{(\beta \, L)}
\end{equation}
where $\beta = k_\bot^2/(4k_p) - \Delta k/2$, and 
$k_\bot$ and  $\Delta k$ now follow from
\begin{align}
& k_\bot^2 \! = k_s^2 \sin^2 \theta_s+k_i^2 \sin^2 \theta_i \, +  \nonumber  \\
&  \qquad \qquad \qquad 2k_s k_i \sin\theta_s \sin \theta_i \cos(\phi_s-\phi_i) \nonumber  \\
&  \mathrm{and}  \quad \Delta k = k_p-k_s \cos \theta_s -k_i \cos\theta_i-k_g \, .
\end{align}

In numerical integration of Eq.~\ref{E:R1} we ignore the $(\theta_{s,i}, \phi_{s,i})$-dependence of $k_{s,i}$ itself, which is a good approximation for propagation near the pump direction.  The theoretical spectra are shown in Fig.~2(b) and, with parameters as in the experiments, the agreement with Fig.~2(a) is excellent.  Theoretical spectra for $\Delta \theta \! > \! 2.5^\circ$ are not shown, but are nearly identical to those for $\Delta \theta \! = \! 2.0^\circ$ and $2.5^\circ$; thus essentially \textit{all} flux is restricted to a cone of $\approx \! 2^\circ$ half-angle.  This compact, beamlike angular emission is a novel property that, in future work, could be essential in obtaining high free-space or fiber-coupled collection efficiencies.  There is also near-symmetry about the degenerate frequency in Fig.~2(b) for $\Delta \theta \! = \! 2.0^\circ$ and $2.5^\circ$.  For a monochromatic pump, this symmetry is a consequence of photon energy conservation if all emitted flux is collected.  For the most part, the symmetry is present in the analogous data of Fig.~2(a), although there is a slight decrease at $\omega_s \! \approx \! 0.8 \! \times \! 10^{15} \, \mathrm{Hz}$ that could arise from light falling outside the collection aperture.  We note that related calculations not presented here predict the signal/idler correlation time to beÊ
only $4.9 \, \mathrm{fs}$ on the optic axis for our experimental parameters ($\lambda_p \! = \! 942.5 \, \mathrm{nm}$ and $\Lambda_g \! = \! 27.4 \, \mu \mathrm{m}$).

In conclusion, we have produced downconverted light having an unprecedented bandwidth of 
$1080 \, \mathrm{nm}$ at a central
wavelength of $1885 \, \mathrm{nm}$. The wideband phase-matching conditions occur for a type 1 process with collinear downconversion at the degenerate frequency 
$\omega_d$ when $k^{(2)}_{s,i} (\omega_d) \! \approx \! 0$.  
A large flux of photons ($3.4\times 10^{11}$ pairs $\mbox{s}^{-1} \mbox{W}^{-1}$) is associated with this large bandwidth, and the angle-dependence of the phase mismatch restricts the light to an easily-collected cone of half-angle $\approx \! 2^\circ$.  The experimental data presented exhibit excellent agreement in direct comparison with theoretical photon number spectra.
Even though our broadband conditions occur only at a particular pump frequency $\omega_p$ for a given crystal, in principle it is possible to employ metamaterials having dispersion designed to produce such effects at any $\omega_p$ desired.  The generation of photon pair states having ultra-short correlation time and an exceptionally high-dimensional spectral entanglement may, in the future, be of key importance for many applications.

\begin{acknowledgements}
We are grateful for discussions with R. S. Cudney, H. F. Alonso, S.
G\"{u}nther, and A. Hemmerich.  This work was supported by Conacyt
grants 46370-F and 49570-F.
\end{acknowledgements}


\end{document}